\documentclass{camera}
\usepackage{graphicx}
\usepackage{amssymb}

\begin{document}

\title{How the \emph{RHESSI} gamma-ray burst
measurements have been affected\\
by the annealing procedure?}

\author{J. \v{R}\'{\i}pa$^{1}$, P. Veres$^{2}$ \and C. Wigger$^{3}$}

\organization{
$^{1}$ Charles University, Faculty of Mathematics and Physics,\\
Astronomical Institute, Prague, Czech Republic\\
$^{2}$ E\"{o}tv\"{o}s Lor\'{a}nd University, Budapest, Hungary\\
$^{3}$ Kantonsschule Wohlen, Switzerland
}

\maketitle

\begin{abstract}
The performance of the nine \emph{RHESSI} germanium detectors has been gradually deteriorating since its launch in 2002
because of radiation damage. To restore its former sensitivity,
the spectrometer underwent an annealing procedure in November 2007. However, it changed the \emph{RHESSI} response and affected
gamma-ray burst measurements, \emph{e.g.}, the hardness ratios and the spectral capabilities below $\sim 100$\,keV.
\end{abstract}

The \emph{RHESSI} satellite \cite{ref01} observes $\simeq 70$ gamma-ray bursts (GRBs) per year \cite{ref02}.
Its energy range is $\simeq 25$\,keV - 17\,MeV, the maximal effective area 200\,cm$^2$.
In November 2007 its spectrometer underwent an annealing procedure to restore its sensitivity
that had been gradually deteriorating because of radiation damage.
This procedure was successful only partly because the
low-energy response was not improved as well as the high-energy
one. As a result the low-energy hardnesses $H21$, measured
after the annealing, are systematically shifted to higher
values (see fig.~1), contrary to the high-energy ratios $H32$,
which remain the same. The $T_{90}$ durations are not
significantly affected. We also compared the spectral
parameters of some GRBs. Whereas the low-energy spectral photon
index of the pre-annealing GRB 061121, measured by \emph{Swift},
Konus, and \emph{RHESSI}, is approximately the~same, for the
post-annealing GRBs 080607 and 080825 our values markedly
differ from the \emph{Swift} or \emph{Fermi} ones (see table~1).

In conclusion, the using of the \emph{RHESSI} data for the future
spectral analyses, over the low energy part of the instrument's
range ($\lesssim 100$\,keV), is problematic without finding a
new instrument's response matrix.
\newline

We acknowledge the support from the grants GAUK 46307, GA\v{C}R 205/08/H005, MSM0021620860, and OTKA K077795.

\begin{figure}
\begin{center}
\includegraphics[width=0.78\textwidth]{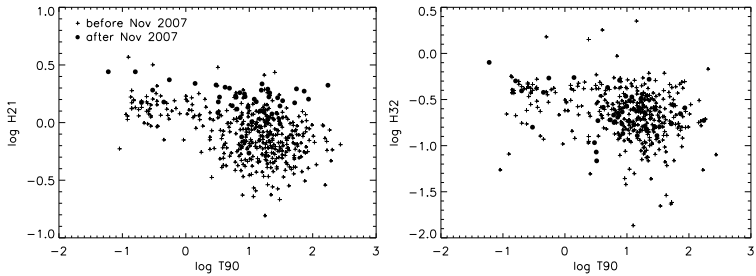}
\caption{Measurements of $H21$
(count ratio between 120-400 and 25-120\,keV)
and $H32$
(ratio between 400-1500 and 120-400\,keV)
before and after the annealing.}
\end{center}
\end{figure}

\begin{center}
\begin{table} \caption{Spectral fits of cutoff power law and Band function of some selected GRBs.
The \emph{RHESSI} off-axis angle for all these GRBs is near right angle.}
\centering
\begin{tabular}{ccccccc}

\hline
\\[-2.5ex]
GRB    &  model & data      &   $\alpha$    &    $\beta$    & $E_p$(keV) & $\chi^2_r$ \\
\hline\hline
       &        & \emph{Swift}     & 1.37$\pm$0.02 &               &            &  1.27      \\
061121 & cpl    & Konus*    & 1.32$\pm$0.05 &               & 606$\pm$80 &  1.01      \\
       &        & \emph{RHESSI}    & 1.37$\pm$0.10 &               & 532$\pm$57 &  1.01      \\
\hline
080607 & cpl    & \emph{Swift}     & 1.15$\pm$0.03 &               &            &  0.70      \\
       &        & \emph{RHESSI}    &\textbf{2.33$\pm$0.18}&       & 432$\pm$19 &  1.39      \\
\hline
080825 & Band   & \emph{Fermi}     & 0.54$\pm$0.21 & 2.29$\pm$0.35 & 180$\pm$23 &  1.23       \\
       &        & \emph{RHESSI}    &\textbf{5.36$\pm$0.86}& 2.92$\pm$0.57 & 256$\pm$25 &  0.80       \\
\hline
\\[-3.5ex]
       &        &           &               &               &            &\tiny*GCN 5837\\
\end{tabular}
\end{table}
\end{center}


\begin{thebibliography}{}
\bibitem[1]{ref01} LIN R. P. ET AL., \emph{SoPh}, \textbf{210} (2002) 3.
\bibitem[2]{ref02} http://grb.web.psi.ch
\end{thebibliography}
\end{document}